\documentstyle[prl,aps,preprint,floats]{revtex}

\newcommand{\half}{\frac{1}{2}}

\input BoxedEPS
\SetRokickiEPSFSpecial %% for dvips by Tom Rokicki, for VMS, Unix
\HideDisplacementBoxes

\begin{document}
%\draft

\pagestyle{myheadings}
\markright{\small  \it Dirac
Particles in Twisted Tubes}
\tighten

\preprint{\vbox{MIT-CTP 2819 \null\hfill\rm January 22, 1999} }

\title{Dirac Particles in Twisted Tubes}

\author{P.~Ouyang, V.~Mohta, and R.L.~Jaffe\\ \null}

\address{Center for Theoretical Physics and Department of Physics \\
Massachusetts Institute of Technology\\ Cambridge, Massachusetts 02139
}

\maketitle
\thispagestyle{empty}

\begin{abstract}

\noindent We consider the dynamics of a relativistic Dirac particle
constrained to move in the interior of a twisted tube by confining
boundary conditions, in the approximation that the curvature of the
tube is small and slowly varying.  In contrast with the
nonrelativistic theory, which predicts that a particle's spin does
not change as the particle propagates along the tube, we find that the
angular momentum eigenstates of a relativistic spin-$\frac{1}{2}$
particle may behave nontrivially.  For example, a particle with its
angular momentum initially polarized in the direction of propagation
may acquire a nonzero component of angular momentum in the opposite
direction on turning through $2\pi$ radians.  Also, the usual
nonrelativistic effective potential acquires an additional factor in
the relativistic theory.
\end{abstract}

%\pacs{}
%\narrowtext
%\newpage

\section{Introduction}

\noindent Twisted tubes, in which particles are confined by an
external potential, have proven to be excellent settings for the study
of quantum mechanics in curved geometries.  They have been studied
extensively in the context of nonrelativistic quantum mechanics, with
particular focus on the case of small and slowly varying
curvature\cite{TT}.
 One important example is the limit of a very narrow tube,
where the ``confining potential'' constrains a particle to move on a
curve, just as a physical bead might be attached to a wire.
Subsequent authors have considered the natural extension to a particle
confined to a small neighborhood of an arbitrary Riemannian manifold
embedded in a higher-dimensional Euclidean space and have applied
their formalism to the energy spectra of molecules\cite{NDIM}. For a
recent survey of the dynamics of a particle in various confining
potentials in a loosely interpreted ``tube'' see Ref.\cite{PM}.
However, analysis has been largely limited to nonrelativistic quantum
mechanics, in which spin does not affect the dynamics; the
relativistic theory has been thoroughly investigated only for
confinement and quantization schemes other than the confining
potential approach\cite{OGAWA}. Here we present for the first time a
treatment of a relativistic Dirac particle in the confining potential
approach using confining boundary conditions.

The phenomena associated with the propagation of a relativistic
fermion within a tube contrast with the simpler and well-known case of
light propagating in an optical fiber, studied in the mid-1980's by
Chiao and Wu\cite{CHIAO}.   They assumed the fiber to have constant
circular cross-section and small and smoothly varying curvature.  In
this approximation, they argued that the two states of circular
polarization (helicity $\pm 1$) propagate along the fiber without
mixing with one another and accumulate opposite Berry phases
proportional to the integral of the torsion along the fiber.  It is
instructive to re-examine their argument that the helicity eigenstates
do not mix\cite{CHIAO}.   First they use the adiabatic theorem to
argue that modes with different transverse waveforms do not mix as
they propagate through a smooth fiber.  The lowest energy level is
doubly degenerate, with total angular momentum component along the
axis of the fiber $\vec J\cdot\hat t =\pm 1$.  The longitudinal mode,
$\vec J\cdot \hat t = 0$, is excluded by gauge invariance.  Following
Chiao and Wu we refer to these modes somewhat loosely as photon states
of helicity $\pm 1$.  In the limit of a thin fiber, the only surviving
terms in the Hamiltonian affecting helicity are of the form $\vec
\kappa\cdot\vec S$, where the ``vector curvature'', $\vec\kappa$, is a
vector in the direction of the principal normal to the fiber and of
length $\kappa$, $1/\kappa$ is the instantaneous radius of curvature
of the fiber, and $\vec S$ is the photon's spin.  Since $\vec
\kappa\cdot\vec S$ is a vector operator, the Wigner-Eckart theorem
allows a change in helicity only by $\Delta h = \pm 1$.  $\Delta h =
\pm 2$ is forbidden, so the helicity states propagate without mixing,
even though they are degenerate -- a consequence of gauge invariance.

The point of departure for our paper is the observation that Chiao and
Wu's argument does not apply to a Dirac fermion, which has doubly
degenerate modes with helicity $\vec J\cdot \hat t = \pm (n +
\frac{1}{2})$.  Once again we refer to these modes somewhat loosely as
fermion states of helicity $\pm (n + \frac{1}{2})$.  In particular,
the ground state modes have $n=0$, so $\vec J\cdot\hat t = \pm \half$.
The Wigner-Eckart theorem does not forbid transitions between these
states because the eigenvalues of $\vec J\cdot\hat t$ differ by 1.  We
cannot exclude the possibility that the degenerate angular momentum
states will mix as they propagate on the basis of symmetry.  Therefore
the Dirac equation demands more careful analysis.

As in Ref.\cite{TT}, we consider a particle confined to the interior
of a tube with circular cross-section.  We work in the rest frame of
the tube.  The tube may be constructed by transporting a circular disc
such that the center of the disc traces a smooth curve ${\vec X}(s)$
and the tangent to the curve is normal to the disc for all values of
the arc-length parameter $s$.  The region swept out by the disc is the
tube.  We model the potential that confines the particle to the tube
by means of a boundary condition.  The boundary condition must
guarantee that the normal component of the probability current $\vec j
= \psi^{\dag} \vec \alpha \psi$ vanishes on the surface of the tube.
In the nonrelativistic problem one can simply impose Dirichlet
boundary conditions.  However, in the Dirac theory, the requirement
that all components of the wavefunction vanish on the boundary is too
stringent, for then in general the Dirac wavefunction will vanish
everywhere.  A less severe condition that will still confine
probability is
\begin{equation}
        -i \vec{\gamma} \cdot \hat n\psi = \psi
        \label{eq1}
\end{equation}
where $\hat n$ is the unit normal (in the outward sense) to the
boundary of the tube\cite{BAG}. That eq.~(\ref{eq1}) guarantees
probability confinement follows from the commutation relations for the
$\gamma$-matrices and the antihermiticity of $\vec{\gamma}$.
Eq.~(\ref{eq1}) reduces to the usual Dirichlet boundary condition on
the ``large'' components of the Dirac spinor in the nonrelativistic
limit.

It is a general feature of the Dirac equation that spin and orbital
angular momentum are not separately conserved.  It is not surprising,
therefore, that the boundary condition, eq.~(\ref{eq1}) does not
commute with the spin operator.  The result is the nontrivial
transport of angular momentum states that we will calculate.  We
could have chosen other boundary conditions, for example the chiral
generalization of eq.~(\ref{eq1}),
\begin{equation}
        -i  \vec{\gamma} \cdot \hat n\psi = e^{i\gamma_{5}\theta}\psi
        \label{eq2}
\end{equation}
parameterized by the chiral angle $\theta$, with a slightly different
phenomenology.  We will explore some of the alternatives in the final
section of this paper.

In Section II we establish coordinates in the interior of the tube
that will be convenient for our purposes.  Section III is a
construction of the dynamical equations of our system in the parallel
transport frame. We write the dynamical equations as those for
propagation in a straight tube with an interaction.  In Section IV we
calculate the necessary basis states for a Dirac particle in a
straight tube.  In Section V we specialize to the approximation that
the curvature is small and slowly varying. This approximation appears
to capture most of the interesting physics. We then expand the
Hamiltonian to lowest nontrivial order in small parameters.  Using the
resulting Hamiltonian, we compute in Section VI the relevant dynamical
equations.  Section VII contains our conclusions and also briefly
discusses possibilities for boundary conditions other than
eq.~(\ref{eq1}).

\section{Elementary Formalism}
We construct a coordinate system for the tube described in Section I
by erecting a right-handed orthonormal frame at each point on the
curve.  We parameterize the curve as ${\vec X}(s)$ where $s$ is the
arc-length parameter.  Since the tangent to the curve is always normal
to the cross-sectional disc that sweeps out the tube, a natural
choice of frame is $\{\hat{e}_1,\hat{e}_2,\hat{e}_3\}$ where
$\hat{e}_1$ and $\hat{e}_2$ are orthogonal unit vectors in the plane
of the disc and $\hat{e}_3$ is the unit tangent to the disc in the
direction of increasing $s$.  For clarity we adopt a summation
convention where Greek indices, $\mu$, $\nu$, etc., are summed over
$1$, $2$, and $3$, whereas Latin indices, $i$, $j$, etc., are summed
only over the ``transverse'' directions $1$ and $2$.  To completely
define the frame, we must specify the $s$-dependence of $\hat{e}_1$
and $\hat{e}_2$.  We require that the frame rotates smoothly as the
disc sweeps out the tube.  That is,
\begin{equation}
        \frac{d\hat{e}_\mu}{ds} = \vec{\omega} \times  \hat{e}_\mu
        \label{eq3}
\end{equation}
where $\vec{\omega}(s)$ is a smoothly varying instantaneous ``angular
velocity'' associated with the rotation of the frame.

Since the curve uniquely determines $\hat{e}_3(s)$, it determines
$\vec{\omega}(s)$ up to the addition of a constant multiple of
$\hat{e}_3(s)$, corresponding to an additional rotation of the
transverse axes about the tangent. For simplicity, we adopt the
``parallel transport'' frame, in which $\vec{\omega}$ has no component
along the tangent. There is a simple visual model of the parallel
transport frame. Imagine a rigid wire with a disc attached to it by a
frictionless sleeve that keeps the disc normal to the sleeve. Without
friction all the torques act in the plane of the disc so the angular
momentum about the tangent direction is conserved. If the disc has no
angular velocity about $\hat{e}_3$ initially, it will not develop any.
In this case, vectors drawn on the disc are said to execute parallel
transport along the wire. A simple calculation shows that the
instantaneous angular velocity of this frame is
\begin{equation}
        \vec{\omega}  =  \hat{e}_3 \times \frac{d\hat{e}_3}{ds}\ .
        \label{eq4}
\end{equation}
A new coordinate system for points inside the tube may now be
specified in terms of the parallel transport frame by
\begin{equation}
        {\vec x(s, \xi^1, \xi^2)} = {\vec X}(s) + \xi^1 \hat{e}_1 +
        \xi^2 \hat{e}_2\ .  \label{eq5}
\end{equation}
In this coordinate system, the spatial metric tensor is diagonal
\begin{equation}
        g_{\mu\nu}= \left( \begin{array}{ccc} 1 &  & \\ & 1 & \\ & &
        (1-\vec{\kappa} \cdot \vec{\xi})^2  \end{array} \right)
        \label{eq6}
\end{equation}
where the vector curvature $\vec{\kappa}$ of ${\bf X}(s)$ is defined
as
\begin{equation}
        \vec{\kappa}\equiv \frac{d\hat{e}_3}{ds} \equiv \kappa^1
\hat{e}_1 + \kappa^2 \hat{e}_2\ .  \label{eq7}
\end{equation}
Comparison with the definition of $\vec\omega$ shows
\begin{eqnarray}
        \vec\omega &=& \hat e_{3}\times\vec \kappa \quad {\rm
        or}\nonumber\\ \omega^{2} &=& \kappa^{1}\quad {\rm and}\quad
        \omega^{1}  = -\kappa^{2}\ .
\end{eqnarray}
The diagonal metric tensor yields a simple gradient operator:
\begin{equation}
        \vec\nabla= \hat{e}_1 \frac{\partial}{\partial \xi^1} +
        \hat{e}_2 \frac{\partial}{\partial \xi^2} + \hat{e}_3
        \frac{1}{1-\vec{\kappa} \cdot \vec{\xi}}
        \frac{\partial}{\partial s}\ .  \label{eq9}
\end{equation}
The volume element for integration in these coordinates is $\sqrt{\det
g_{jk}} ds d^2 \xi = (1-\vec{\kappa} \cdot \vec{\xi}) ds d^2
\xi$. Since the tube intersects itself if $\vec{\kappa} \cdot
\vec{\xi}$ equals one, we require that $\kappa R < 1$ where $R$ is the
radius of the disc.

An important subtlety is that while the choice of the parallel
transport frame is always valid locally, it is not always possible to
adopt it globally.  In particular, if the curve is closed, parallel
transport of the frame through one circuit of the curve need not
return the frame to its original orientation.  This geometric feature
manifests itself in classical physics as the well-known Hannay's
angle\cite{Hannay,TT}. For the purposes of this paper, we assume that
our tube continues indefinitely without intersecting itself.

\section{Dirac Equation in the Parallel Transport Frame}

In free space, with units such that $\hbar = c = 1$, the Hamiltonian
in the Dirac equation ${\cal H} \Psi = E \Psi$ is
\begin{equation}
        {\cal H} = (-i \vec{\alpha} \cdot \vec\nabla + \beta m + V)\ .
        \label{eq3.1}
\end{equation}
We define $\alpha$ and $\beta$ as usual by  $\alpha^i = \gamma^0
\gamma^i$ and $\beta = \gamma^0$; in the chiral basis that we employ,
\begin{equation}
        \gamma^0 = \pmatrix {0 & 1 \cr 1 & 0} , \,\, \gamma^{i} =
        \pmatrix{0&\sigma^{i} \cr -\sigma^{i} & 0}\,\,,\gamma_5 =
        i\gamma^0 \gamma^1 \gamma^2 \gamma^3 = \pmatrix {-1&0\cr 0
        &1}\ .  \label{eq3.2}
\end{equation}

In the coordinate system defined in Section II, the Dirac
Hamiltonian takes the form
\begin{equation}
        {\cal H} = -i \vec{\alpha} \cdot \hat{e}_j
        \frac{\partial}{\partial \xi^j} - \frac{i}{\Delta}
        \vec{\alpha} \cdot \hat{e}_3 \frac{\partial}{\partial s} +
        \beta m \label{eq3.3}
\end{equation}
where $\Delta = 1-\vec{\kappa} \cdot \vec{\xi}$.  We normalize our
wavefunctions $\Psi$ so that
\begin{equation}
        \int d^{3}x \,\Psi^{\dag}\Psi = \int ds\, d^2\xi\, \Delta
        \Psi^{\dag} \Psi = 1\ .  \label{eq3.4}
\end{equation}
The factor of $\Delta$ in the integration measure complicates
calculations and questions of hermiticity.  Therefore we define new
wavefunctions $\psi$ by
\begin{equation}
        \psi = \Delta^{\frac{1}{2}}\Psi \label{eq3.5}
\end{equation}
which have the normalization $\int ds d^2 \xi \,\psi^{\dag} \psi = 1$.
The original Dirac equation can be recast in terms of a Hamiltonian
operator $\tilde{H}$, defined by $\tilde{H} = \Delta^{\frac{1}{2}}
{\cal H} \Delta^{-\frac{1}{2}}$, which acts on the wavefunctions
$\psi$,
\begin{equation}
        \tilde{H} = -i \vec{\alpha} \cdot \hat{e}_j
        \frac{\partial}{\partial \xi^j} - \frac{i}{2 \Delta}
        \vec{\alpha} \cdot \vec{\kappa} - \frac{i}{\Delta}
        \vec{\alpha} \cdot \hat{e}_3 \Bigl( \frac{\partial}{\partial
        s} + \frac{1}{2 \Delta} \frac{\partial (
        \vec{\kappa}\cdot \vec{\xi} )}{\partial s} \Bigr) +
        \beta m\ .  \label{eq3.6}
\end{equation}

This Hamiltonian is difficult to calculate with because the matrices
$\vec\alpha\cdot\hat e_{\mu}$ are $s$-dependent. That is, the momentum
operators are expressed with respect to the rotating coordinate frame
while the spin operators are expressed with respect to some fixed
rectilinear frame. To remedy this problem, we change the spin
basis. We define a unitary transformation $\Omega$ by
\begin{equation}
       \Omega(s) \vec{\alpha} \cdot \hat{e}_\mu(s)\Omega^{\dag}(s)=
\alpha_{\mu} \label{eq3.7}
\end{equation}
for all $s$.  Here we have made the $s$ dependence of the unit vectors
$\hat e_{\mu}$ explicit.

Differentiating the above condition with respect to $s$ and using
eq.~(\ref{eq3}) for $d\hat e_{\mu}/ds$, we find
\begin{equation}
        \frac{d\Omega}{ds} = \frac{i}{2} \vec{\Sigma} \cdot
\vec{\omega} \,\Omega \label{eq3.8}
\end{equation}
where $\Sigma^i$ are the $4 \times 4$ Dirac spin matrices. Therefore,
\begin{equation}
        \Omega(s) = {\cal P} \exp \left[ \frac{i}{2} \int^s ds' \,
\vec{\Sigma} \cdot \vec{\omega}(s') \right] \label{eq3.9}
\end{equation}
where ${\cal P} $ denotes path-ordering along the tube.

To take advantage of this transformation, we once again define new
wavefunctions,
\begin{equation}
        \chi (s,\vec \xi)= \Omega (s)\psi(s,\vec \xi) \label{eq3.10}
\end{equation}
and a new Dirac Hamiltonian, $H$,
\begin{equation}
        H =  \Omega \tilde{H} \Omega^{\dag}\ .  \label{eq3.11}
\end{equation}
The form of eq.~(\ref{eq3.9}) indicates that the transformation
$\Omega(s)$ just represents a change of spin basis to the spinors
representing spin up and down in the $\hat{e}_3(s)$
direction. Although this transformation replaces the $\alpha$-matrices
as intended, it generates a ``gauge term'' in $H$,
\begin{equation}
        \frac{-i}{\Delta} \, \Omega \, \vec{\alpha} \cdot \hat{e}_3
        \frac{\partial \Omega^{\dag}}{\partial s} = -\frac{1}{\Delta}
        \alpha^3 \Bigl( \frac{1}{2} \vec{\Sigma} \cdot \vec{\omega}
        \Bigr)\ .  \label{eq3.12}
\end{equation}
Simplifying $\alpha^3 \, \vec{\Sigma} \cdot \vec{\omega}$ yields $-i
\vec{\alpha} \cdot \vec{\kappa}$.  Thus, this gauge term cancels the
second term in eq.~(\ref{eq3.6}) giving the following relatively
simple Hamiltonian for $\chi$:
\begin{equation}
        H=-i {\alpha}^j \frac{\partial}{\partial \xi^j} -i {\alpha}^3
        \frac{\partial}{\partial s} + \beta m - i {\alpha}^3 \Bigl(
        \frac{\vec{\kappa}\cdot
        \vec{\xi}}{\Delta}\frac{\partial}{\partial s} + \frac{1}{2
        \Delta} \frac{\partial (\vec{\kappa}\cdot
        \vec{\xi})}{\partial s} \Bigr)\ . \label{eq3.13}
\end{equation}
The transformations we have performed reduce our problem to the study
of this perturbed Dirac Hamiltonian in a right circular cylinder,
$(\xi^{1})^{2}+(\xi^{2})^{2} \le R^{2}$, $-\infty\!<\!s\!<\!\infty$,
with respect to a fixed rectilinear frame.  The price of this
simplification is the variety of curvature dependent terms in
eq.~(\ref{eq3.13}).

\section{Decomposition into Transverse Eigenstates}

 We look for eigenstates of the Hamiltonian in eq.~(\ref{eq3.13}) of
the form
\begin{equation}
        \chi(s,\vec\xi) = \sum_{\{\rho\}}c_{\{\rho\}}(s)
        \eta_{\{\rho\}}(\vec\xi)  \label{eq4.1}
\end{equation}
where the $\{\eta_{\{\rho\}}\}$ are eigenstates of the transverse
Hamiltonian,
\begin{equation}
        H_{\bot}= -i \alpha^j \frac{\partial}{\partial \xi^j} + \beta
        m \label{eq4.2}
\end{equation}
satisfying the boundary condition
\begin{equation}
        \left.-i\vec\gamma\cdot\hat r\eta = \eta \right|_{\rm r=R}\ .
\label{eq4.3}
\end{equation}
Here we have introduced plane polar coordinates, $r=|\vec\xi|$ and
$\varphi=\tan^{-1}\xi^{2}/\xi^{1}$.  The labels $\{\rho\}$ denote the
quantum numbers necessary to specify the eigenstates of $H_{\perp}$
completely.

$H_{\bot}$ is just the free Dirac Hamiltonian in one lower dimension.
$S_3$ and $L_3$, the tangential components of the spin and orbital
angular momentum respectively, do not separately commute with
$H_{\perp}$ but $J_3 = L_3 + S_3$ does. $J_3$ also commutes with the
operator $\vec{\gamma} \cdot \hat{r}$ in the boundary condition.  In
addition, the operator $\gamma^{0} S_3$ commutes with $H_{\perp}$,
$J_3$, and $\vec{\gamma} \cdot \hat{r}$.  We define the ``transverse
energy'' as the eigenvalue $\lambda$ of $H_{\bot}$.  An eigenstate of
$H_{\bot}$ with eigenvalue $\lambda$ may thus be chosen to be an
eigenstate of the helicity operator $J_3$ with eigenvalue $\nu$ ($\nu
= n - 1/2$ for some integer $n$) and of the operator $\gamma^{0} S_3$
with eigenvalue $\varepsilon$ ($\varepsilon = \pm \half$). For
simplicity, we use the notation $\{ \nu^{\pm}, \lambda \}$ when
specifying these quantum numbers.

\begin{figure}[ht]
        $$\BoxedEPSF{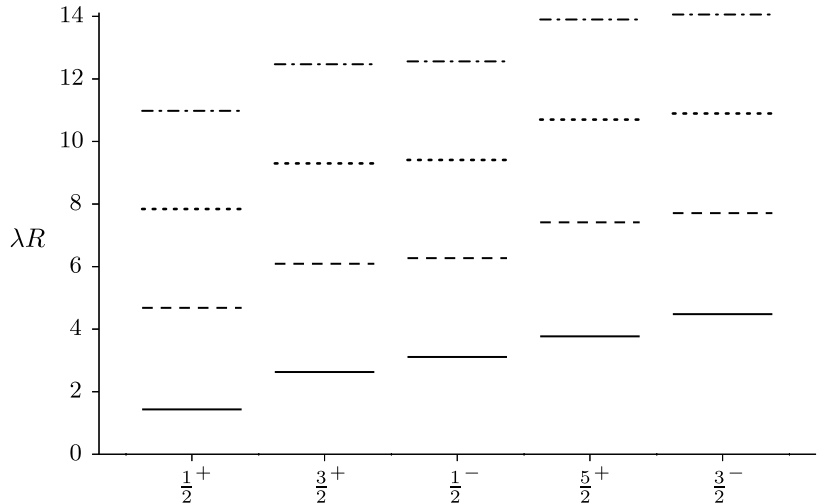}$$ \caption{Summary of
        the low-lying transverse energy eigenstates in the extreme
        relativistic limit, $m\to 0$.  Only positive $\lambda$ are
        shown.  Each level is twofold degenerate corresponding to
        $\nu^{+}\leftrightarrow-\nu^{-}$.  Radial excitations are 
	displayed with dashed lines to correlate with Fig.~\ref{fig3}.}
\label{fig1}
\end{figure}

$H_{\perp}$ has the usual Dirac charge conjugation symmetry, so each
eigenstate with eigenvalue $+|\lambda|$ is paired with another with
eigenvalue $-|\lambda|$.  Since $\alpha^3$ anticommutes with
$H_{\perp}$ and  $\gamma^0 S_3$  and commutes with $J_3$ and
$\vec{\gamma} \cdot \hat{r}$, $\alpha^3 \eta_{\nu^{\pm}, \lambda} =
\eta_{\nu^{\mp}, - \lambda}$.  The transformation that sends $J_{3}$
to $-J_{3}$ commutes with $H_{\perp}$ and $\vec{\gamma} \cdot
\hat{r}$ and anticommutes with $\gamma^0 S_3$.  Thus, the states
with $\nu^{+}$ have the same eigenvalue $\lambda$ as states with
$-\nu^{-}$.  For each value of $\nu$ and $\varepsilon$, there are
towers of radial excitations, one for positive $\lambda$ and one for
negative $\lambda$, obtained by adding more nodes to the radial
wavefunction.  A summary of the eigenstates of $H_{\perp}$ for small
$\nu$ and $\lambda$ is given in Fig.~\ref{fig1}.

With these general considerations in mind, we construct the explicit
representation of the wavefunction $\eta_{\nu^{\pm},\lambda}$ for
positive $\lambda$:
\begin{equation}
        \eta_{\nu^{\pm},\lambda}(r,\varphi) =
        N_{\nu^{\pm},\lambda}\pmatrix{\zeta^{\pm 1}  J_{n-1}(qr)
        e^{i(n-1) \varphi}\cr -i J_n(qr) e^{in \varphi}\cr  \pm
        \zeta^{\pm 1} J_{n-1}(qr) e^{i(n-1) \varphi}\cr \pm i  J_n(qr)
        e^{in \varphi}} \label{eq4.4}
\end{equation}
where $n$ is an integer with $\nu = n - \half$, $J_n$ is a Bessel
function of the first kind, $q = + \sqrt{{\lambda}^2 - m^2}$, $\zeta
\equiv \frac{m+\sqrt{q^2+m^2}}{q} $, and $q$ satisfies the following
equation obtained from the boundary condition:
\begin{equation}
        J_n(qR) = \pm \zeta^{\pm 1} J_{n-1}(qR)\ .  \label{eq4.5}
\end{equation}
The normalization $N$ is fixed up to a phase by the condition
\begin{equation}
        \int d^2\xi \eta^{\dag} \eta = 1\ .  \label{eq4.6}
\end{equation}

The unusual form of the spinors in the nonrelativistic limit is due
to our choice of the chiral representation of the Dirac matrices
($\gamma_{5}= {\rm diag} (-1,-1,1,1)$).  Transforming the eigenstates
to the Dirac basis ($\gamma_{0}= {\rm diag} (1,1,-1,-1)$), we find
that only one of the two upper components and one of the two lower
components of each wavefunction are nonzero.  In the nonrelativistic
limit, for the states with positive $\lambda$, the lower components in
the Dirac basis tend to zero.  These are eigenstates of $S_3$ {\it
and\/} $L_3$, in agreement with the fact that $\gamma^0 S_3$ is just
$S_3$ on the upper components in the Dirac basis.  We use the chiral
basis because the wavefunctions can be expressed more compactly away
from the nonrelativistic limit and because computation of matrix
elements is simpler.

For a straight tube, the energy eigenstates with helicity $\nu$
and longitudinal momentum $k$ are obtained from the ansatz,
eq.~(\ref{eq4.1}):
\begin{equation}
        \chi(s,\vec\xi) = \exp{iks}
        \Bigl(\eta_{\nu^{\pm},\lambda}(\vec\xi) +
        \frac{k\, \hbox{sgn}(\lambda)}
        {\sqrt{\lambda^2 + k^2} +
        |\lambda|}
        \eta_{\nu^{\mp},-\lambda}(\vec\xi)\Bigr)\ .
        \label{eq4.6a}
\end{equation}
In a twisting tube, the energy eigenstates could be superpositions of
several $\{\eta_{\nu^{\pm},\lambda}\}$ with nontrivial
$\{c_{\nu^{\pm},\lambda}(s)\}$ determined by eq.~(\ref{eq3.13}).

In later sections, we will study the propagation and mixing of the
ground states of the transverse Hamiltonian, i.e., those with the
smallest possible value ($\lambda_{0}$) for $|\lambda|$.  For
concreteness, we present the explicit form of the wavefunctions for
this fourfold degenerate set of states:
\begin{eqnarray}
    \eta_1 \equiv \eta_{\half^{+},\lambda_{0}}&=& N \pmatrix { \ \zeta
    J_{0}(qr) \cr -i J_1(qr) e^{i \varphi}\cr \zeta J_0(qr) \cr i
    J_1(qr) e^{i \varphi} } \qquad \eta_3 \equiv
    \eta_{\half^{-},-\lambda_{0}}= N \pmatrix { \zeta J_{0}(qr) \cr i
    J_1(qr) e^{i \varphi}\cr- \zeta J_0(qr) \cr i J_1(qr) e^{i
    \varphi} }\nonumber\\[2ex]
    \eta_2 \equiv \eta_{-\half^{-},\lambda_{0}}&=& N \pmatrix{ -i
    J_1(qr) e^{-i \varphi} \cr \zeta J_{0}(qr) \cr i J_1(qr) e^{-i
    \varphi}\cr \zeta J_{0}(qr) } \qquad \eta_4 \equiv
    -\eta_{-\half^{+},-\lambda_{0}}= N \pmatrix{ i J_1(qr) e^{-i
    \varphi}  \cr \zeta J_{0}(qr) \cr i J_1(qr) e^{-i \varphi}\cr
    -\zeta  J_{0}(qr) } \label{eq4.7}
\end{eqnarray}
where we have moved some of the factors in the representation in
eq.~(\ref{eq4.4}) into the normalization for later convenience and
used $J_{-n} = (-1)^n J_n$. The eigenvalue condition
\begin{equation}
        J_0(qR) = \frac{1}{\zeta} J_1(qR) \label{eq4.8}
\end{equation}
determines $\lambda_{0}$, and the normalization condition
eq.~(\ref{eq4.6}) then fixes $N$:
\begin{equation}
        \frac{1}{N^2} = 2 \pi R^2 \Bigl( \frac{(\zeta^2+1)^2}{\zeta^2}
        -  \frac{2}{\zeta qR} \Bigr) J_1^2(qR)\ .
\end{equation}
In the nonrelativistic limit ($m \gg q$), eq.~(\ref{eq4.8}) reduces
to $J_0(qR) = 0$, a familiar result, and in the ultrarelativistic
limit ($m \rightarrow 0$) it becomes $J_0(qR) = J_1(qR)$.

\section{Adiabatic Condition}

In addition to assuming that the vector curvature, $\vec{\kappa}$, is
a smooth function of $s$, we assume that it is nonzero only in a
finite region. Thus
the tube consists of two infinitely long straight regions joined by a
curved region of finite length.  Our goal is to propagate the
unperturbed solutions to the Dirac equation, labelled by their
longitudinal momentum, $k$, and their transverse quantum numbers, $\{
\nu^{\pm}, \lambda \}$, through the region of nonzero curvature.
We expect both scattering states ($k$ real) and bound states
($k$ imaginary).  By the ``transverse quantum
numbers'' of a tube eigenstate, we mean the quantum numbers
associated with the {\it first term \/} of its representation in the
form given by eq.~(\ref{eq4.6a}).

We restrict our analysis to the case of adiabatic propagation in a
mildly curved tube that dominates the work in this field. This
amounts to the conditions
\begin{eqnarray}
\left| \kappa R \right| &\ll
&1  \nonumber\\[2ex] \left| \frac{\partial \kappa^j}{\partial s} R^2
\right| &\ll &1\ .
\label{eq5.0}
\end{eqnarray}
In this approximation to the nonrelativistic problem, states with
different transverse waveforms do not mix\cite{LIN}. Because of the
analogy between time development with a slowly varying $V(t)$ and
$s$-development with a slowly varying curvature, this is known as the
``adiabatic approximation''.  The generalization of the adiabatic
approximation to the Dirac case is that only states with the same
values of $|\lambda|$ can mix when propagating in a smooth enough
tube.  For example, the four states with $|\nu|=\half$ given
explicitly in eq.~(\ref{eq4.7}) can mix adiabatically.  Since this
subspace includes states of different $\vec J\cdot\hat e_{3}$,
interesting angular momentum transport is still possible in the
adiabatic limit.

For simplicity, we consider the case of small longitudinal momentum.
We assume that the square of the longitudinal momentum of a state
under consideration is much smaller than the difference between the
square of its transverse energy and that of other states with
differing transverse energy.  For a particular state with a set of
transverse quantum numbers $\rho$ , this amounts to the condition
\begin{eqnarray}
\left| k R \right| &\ll
&d_{\rho}
\label{eq5.1}
\end{eqnarray}
where $d_{\rho} =\rm min_{\rho'} \sqrt{|(q_{\rho'} R)^2 - (q_{\rho}
R)^2|}$ and $\rho'$ runs over all sets of transverse quantum numbers
such that $q_{\rho'} \neq q_{\rho}$.

\begin{figure}[ht]
        $$\BoxedEPSF{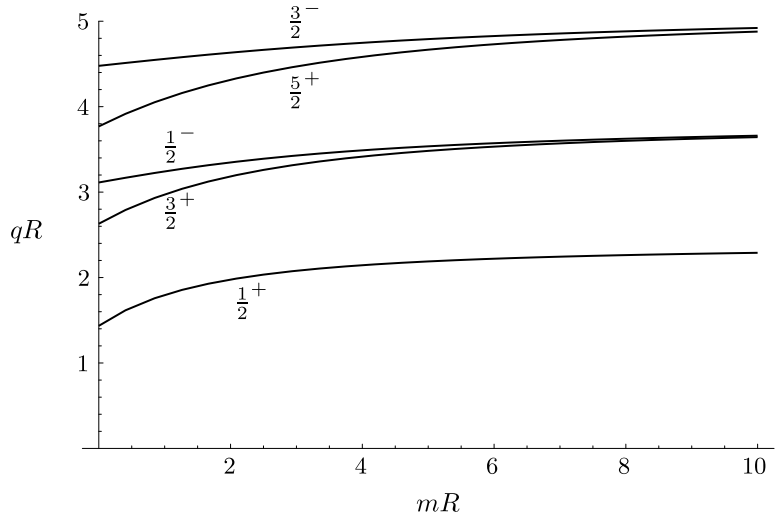}$$ \caption{Dependence of $qR$ on $mR$ for
        the five lowest levels in Fig.~\ref{fig1}} \label{fig2}
\end{figure}

For a physical tube, these conditions are easily satisfied.  For a
fixed mass $m$ and a given curve with smoothly varying vector
curvature $\vec{\kappa}(s)$ which is nonvanishing only in a finite
region, we can always choose a nonzero $R$ small enough to satisfy
each of the above conditions.  There are, however, some subtleties
associated with this condition.  For a fixed tube radius $R$, making
$m$ arbitrarily large makes $d_{\rho}$ arbitrarily small for most
$\rho$.  The additional degeneracies that develop as
$mR\to\infty$ originate in the decoupling of spin and orbital angular
momentum.  Thus, pairs of states with $L_{3}=1$ and $S_{3}=\pm\half$,
which are split by relativistic effects, become degenerate as
$mR\to\infty$ as shown in Fig.~(\ref{fig2}).  The cases
$\rho=\{\frac{1}{2}^{+},\lambda\}$ or $\{-\frac{1}{2}^{-},\lambda\}$
are an exception.  These states, which have $L_{3}=0$ and $S_{3}=\pm
\half$ are degenerate for all $mR$, and no new degeneracies arise as
$mR\to\infty$.  Again, see Fig.~(\ref{fig2}).

Of course, this is not problematic because we know exactly how spin
and orbital angular momentum eigenstates behave in the
nonrelativistic limit: the Dirac problem reduces to the scalar
problem solved, for example, in Ref.\cite{TT}, accompanied by a
dynamically trivial spin label.

Each of the conditions in eqs.~(\ref{eq5.0}) and (\ref{eq5.1}) has a
small parameter associated with it. We let $\delta$ denote the largest
of these parameters.
To simplify the analysis of the relative importance of
terms in the Hamiltonian, we introduce scaled coordinates $u^j \equiv
\xi^j/R$ and multiply the Dirac equation by $R$ to obtain $H'\chi = E'
\chi$ where $H' = R H$ and $E'= R E$ are the scaled Hamiltonian and
energy respectively.  From eq.~(\ref{eq3.13}), we obtain
\begin{equation}
        H'=-i \alpha^j \frac{\partial}{\partial u^j} + \beta m R - i
        {\alpha}^3 R \frac{\partial}{\partial s} + \Bigl( \frac{R
        \vec{\kappa}\cdot\vec{u}}{\Delta} R \frac{\partial}{\partial
        s} + \frac{R}{2\Delta^2} \frac{\partial \left(R
        \vec{\kappa}\cdot \vec{u}\right)}{\partial s}
        \Bigr)\ . \label{eq5.1a}
\end{equation}

So far, this equation is exact. The first two terms are zeroth order
in $\delta$, the third term is first order in $\delta$, and the last
two terms are second order in $\delta$. To second order in $\delta$,
we can replace ${\Delta}^{-1}$ by $1$:
\begin{equation}
         H' = -i \alpha^j \frac{\partial}{\partial u^j} + \beta m R -
         i \alpha^3 \Bigl( \left(1 + R \vec{\kappa} \cdot
         \vec{u}\right) R \frac{\partial}{\partial s} + \frac{R}{2}
         \frac{\partial \left(R \vec{\kappa}\cdot
         \vec{u}\right)}{\partial s} \Bigr) + {\cal
         O}({\delta}^3)\ . \label{eq5.1b}
\end{equation}
For later use, we also compute $H'^2$ to second order in $\delta$:
\begin{equation}
        H'^2 = -\nabla_{u}^2 + R^2 m^2 - R^2
        \frac{\partial^2}{\partial s^2} - i R^2 \vec{\Sigma} \cdot
        \vec{\omega} \frac{\partial}{\partial s} - \frac{R^2}{2}
        \frac{\partial}{\partial s} \left(\vec{\Sigma} \cdot
        \vec{\omega} \right) + {\cal O}({\delta}^3)\ .  \label{eq5.2a}
\end{equation}
Here $\vec\Sigma$ is the Dirac spin matrix, $\vec \Sigma =
\pmatrix{\vec\sigma & 0 \cr 0 & \vec\sigma}$.

Having systematically expanded the Hamiltonian to second order in
small quantities, we may now return to unscaled coordinates.  The
adiabatic condition implies that the effective potential mixes two
transverse eigenstates $\eta_\rho$ and $\eta_{\rho'}$ to second order
in $\delta$ only if $|\lambda_{\rho}|$ = $|\lambda_{\rho'}|$, or
equivalently if $q_{\rho} = q_{\rho'}$.  Thus, to second order in
$\delta$, the mixing amplitudes outside of the four dimensional
subspaces of fixed $q$ are negligible.

As is often the case with the Dirac equation, it is easier to work
with the square of the Dirac operator instead of the first order form.
Since the curvature vanishes for $s \leq s_{0}$ for some $s_0$, the
solution $\chi(s,\vec\xi)$ is given by the straight cylinder problem
for $s<s_{0}$.  This determines both $\chi(s_{0},\vec\xi)$ and
$d\chi(s,\vec\xi)/ds|_{s=s_{0}}$.  To solve the Dirac equation to the
desired order, we project the equation onto the four-dimensional
subspace containing $\chi(s_{0},\vec\xi)$ to obtain four coupled first
order equations.  Equivalently, any solution of the projected equation
must also satisfy the projection of the equation $H^2 \chi = E^2 \chi$
to second order in $\delta$.  The initial conditions on $\chi$ and
$d\chi/ds$ at $s=s_{0}$ determine a unique solution of the second
order equation, which must also be the unique solution of the first
order equation.  From now on, we study the second order equation,
eq.~(\ref{eq5.2a}). We complete the square and divide by $R^2$, to
obtain
\begin{equation}
        H^2 = - \nabla_{\xi}^2 + m^2 -\Bigl(\frac{\partial}{\partial
        s} -\frac{i}{2}\vec{\Sigma} \cdot \vec{\omega} \Bigr)^2  -
        \frac{\kappa^2}{4} + {\cal 
        O}(\delta^{3}/R^{2})\ .
\label{eq5.2}
\end{equation}
Notice that the terms of order $\delta^{3}/R^{2}$, can be neglected in
the adiabatic approximation.  $H^{2}$ is simpler to study than $H$
because it only mixes states with the same $\lambda$.  Thus the mixing
problem has been reduced from a $4\times 4$ to a $ 2\times 2$ problem.
For simplicity, we refer to the two dimensional space of states (with
$\vec J\cdot\hat t = \pm \nu$) that can mix under $H^{2}$ as the
``adiabatic subspace''.  [This simplification is analogous to the
Foldy-Wouthuysen transformation that eliminates the lower components
of the Dirac spinor in the nonrelativistic limit.]  This remarkably
simple result is interpreted in the following section.

\section{Propagating Modes and Spin Evolution}

This Hamiltonian (squared) of eq.~(\ref{eq5.2a}) appears to be the
same one that one would obtain in a nonrelativistic theory.  The
curvature induced ``potential", $-\kappa^{2}/4$, is well known.  The
``gauge term'', $-\frac{i}{2}\vec{\Sigma} \cdot \vec{\omega}$, looks
like it undoes the rotation, $\Omega(s)$, that put us into the moving
frame.  If it did, a stationary experimenter outside the tube would
observe a spin vector that always points in the same direction as the
particle propagates in $s$.  However, this description of propagation
by this Hamiltonian must not be correct: Because $\vec\Sigma \cdot
\vec\omega$ fails to commute with the boundary condition
eq.~(\ref{eq1}), it cannot in general be diagonalized with respect to
the transverse basis states identified in Section IV.

To obtain the correct physical description, we must remember that
$H^{2}$ is to be projected onto the adiabatic subspace containing two
states of opposite helicity ($\nu=\pm\half$).  The externally
prescribed vector curvature, $\vec\kappa$, explicitly breaks rotation
invariance and allows states of different helicity to mix.  However,
the symmetry violating terms in $H^{2}$ are at most vector operators
(note $(\vec\Sigma\cdot\vec\omega)^{2}=\kappa^{2}$), and by the
Wigner-Eckart theorem can only induce transitions between states with
helicity differing by unity.  Thus {\it mixing of different helicity
states can only occur if $|\nu|= \half$.} For all the states with
$|\nu|>\half$, helicity, and in fact all transverse quantum numbers,
are conserved.  This is exactly analogous to the propagation of
photons discussed by Chiao and Wu\cite{CHIAO}.

States with $|\nu|=\half$ mix as they propagate.  The crucial feature
that distinguishes the relativistic and nonrelativistic situations is
that {\it the states with $\nu = \pm\half$ are not eigenstates of
$\Sigma_{3}$.} Instead they are eigenstates of $J_{3}$ and $\gamma^0
S_3$. Since the ground states do not acquire any additional
degeneracies in the nonrelativistic limit, we can apply our formalism
to them in this limit.  The ground states become $L_{3}=0$ and $S_3=
\pm \half$ states. The gauge term in eq.~(\ref{eq5.2}) exactly undoes
the rotation $\Omega(s)$ and leads to trivial spin transport along the
tube. However, in the relativistic case, the matrix elements of
$\vec\Sigma$ between states with $\nu=\pm\half$ are reduced by a
``depolarizing factor'' which reflects the mixing of spin and orbital
angular momentum.

We study the spin transport in the interesting case of
$\nu=\pm\half$. For concreteness, we calculate the matrix elements of
the operator $\vec{\Sigma} \cdot \vec{\omega}$ in the {\it lowest \/}
adiabatic subspace, $\{\eta_{j}\}$, enumerated in eq.~(\ref{eq4.7})
\begin{equation}
        \langle j | \vec{\Sigma} \cdot \vec{\omega} | k \rangle = \int
        d^2 \xi\, \eta^{\dag}_{j}\, \vec{\Sigma} \cdot
        \vec{\omega}\, \eta_{k}. \label{eq5.3}
\end{equation}
For positive $\lambda$, only the (12) and (21) matrix elements are
nonzero, and for negative $\lambda$, only (34) and (43) matrix
elements
are nonzero,
\begin{eqnarray}
        \int d^2 \xi\,  \eta^{\dag}_{1}\, \vec{\Sigma} \cdot
        \vec{\omega}\, \eta_{2} &=& \int d^2 \xi\,  \eta^{\dag}_{3}\,
        \vec{\Sigma} \cdot \vec{\omega}\, \eta_{4} \nonumber\\ & = &
        2N^2 (\omega^1 -i \omega^2) \zeta^2 \int_0^R \int_0^{2 \pi} r
        dr\,  d\varphi\,  J_0(qr)^2 \nonumber\\ & = & 2 \pi N^2 R^2
        (\omega^1 -i \omega^2) (\zeta^2 + 1) J_1(qR)^2\nonumber\\ &=&
        (\omega^1 -i \omega^2) D \label{eq5.4}
\end{eqnarray}
where we have defined the ``depolarizing factor'' $D$ as
\begin{equation}
D \equiv \frac{\zeta^2 (\zeta^2 + 1)}{(\zeta^2 + 1)^2 - \frac{2
\zeta}{qR}}.
\label{eq5.5}
\end{equation}
A similar expression holds for each excited subspace with
$|\nu|=\half$.  For a subspace containing a state in the tower over
$\eta_{\half^{+}, \lambda_0}$, one just changes $q$ accordingly in the
expression for $D$.  For a subspace containing a state
$\eta_{\half^{-}, \lambda}$, one needs to change $q$ and replace
$\zeta$ by $-\zeta^{-1}$, in the expression for $D$.  The general
expression is
\begin{equation}
D(q, \pm) \equiv \frac{\zeta^{\pm 2} (\zeta^{\pm 2} + 1)}{(\zeta^{\pm
2} + 1)^2 \mp \frac{2 \zeta^{\pm 1}}{qR}}
\label{eq5.8}
\end{equation}
where the sign $\pm$ is the product of the signs of the transverse
quantum numbers and is the same for the 4 basis states in each
adiabatic subspace. An important caveat is that this formula fails
to hold in the nonrelativistic limit for the lower choice of sign.
The states with $\nu = \frac{1}{2}$ and those with $\nu = \frac{3}{2}$
both reduce to $L=1$ and have the same transverse energy in the
nonrelativistic limit; the additional degeneracies signal that these
subspaces do not evolve adiabatically.  However, as remarked earlier,
we know that in this limit $D=1$ and a straightforward calculation
extending to a 16-dimensional subspace of states verifies this fact.

The factor $(\omega^1 -i \omega^2)$ in eq.~(\ref{eq5.4}) shows the
parallel between the matrix \emph{elements} of $\vec\Sigma \cdot
\vec\omega$ in the $\{\eta_{j}\}$ basis and its matrix in the basis of
eigenstates of $\Sigma_3$.  The constant of proportionality, $D$,
measures the rate at which the spin vector rotates in the parallel
transport frame as the wave propagates.  In the nonrelativistic
limit, $\zeta \to \infty$ so $D \to 1$, and the spin completely
decouples from the evolution of the tube.  If $D$ were zero for some
choice of parameters, the spin would remain fixed in the parallel
transport frame as for the photon.  In the ultrarelativistic limit,
$\zeta = 1$ and $qR \simeq 1.434$, the lowest root of the equation
$J_0(qR)= J_1(qR)$.  Substitution gives $D \simeq 0.767$, so that even
in the extreme relativistic limit helicity is not conserved.

We now project the second-order Dirac equation, $H^2 \chi = E^2 \chi$,
onto the lowest multiplet, $\{\eta_{j}\}$.  Writing
$\chi=\sum_{j=1}^{4} c_{j}(s) \eta_{j}(\vec\xi)$, the $4 \times 4$
matrix equation for the four component vector $C(s) =
(c_{1},c_{2},c_{3},c_{4})$ is
\begin{equation}
 \Bigl[ -\Bigl(\frac{\partial}{\partial s} -\frac{i}{2}D \vec{T} \cdot
\vec{\omega} \Bigr)^2 + q^{2}- \frac{\kappa^2 D^2}{4}+ m^2\Bigr]
C(s) = E^{2} C(s)
\label{eq5.7}
\end{equation}
where $\vec{T}$ is the vector of ($4\times 4$) Pauli matrices, $\vec T
= \pmatrix{\vec\tau & 0 \cr 0 & \vec\tau}$, that generates the
rotations of the $J_{3}=\pm\half$ eigenstates.  Just as the Pauli
matrices, $S_j$, rotate $S_3=1/2$ states into $S_3=-1/2$ states, the
matrices $T_j$ rotate the $J_3=1/2$ states into $J_3=-1/2$ states.

Eq.~(\ref{eq5.7}) is our fundamental result.  Relativistic effects
appear in two places.  First, as pointed out earlier, the rotation
back to the laboratory fixed frame is only complete in the
nonrelativistic limit, when $D=1$.  As $q$ increases relative to $m$,
$D$ decreases. The ideal case of $D=0$, in which the helicity is
conserved, is not reached even in the extreme relativistic limit. The
following section discusses opportunities presented by other choices
of boundary conditions. Second, the depolarizing factor appears in the
effective scalar potential, $-\kappa^{2}D^{2}/4$.  If we apply the
obvious unitary transformation $C_D(s)= \Omega^{\dag}_D(s) C(s)$ (see
eq.~(\ref{eq3.9}) ), where
\begin{equation}
\Omega_D (s) \equiv  {\cal P} \exp \left[ -\frac{i}{2} D \int^s ds'\, 
\vec{T} \cdot \vec{\omega}(s') \right],
\label{eq5.9}
\end{equation}
then the Dirac equation for $C_D(s)$ reduces to the scalar case
\begin{equation}
\Bigl( -\frac{\partial^2}{\partial s^2} - \frac{\kappa^2 D^2}{4}
\Bigr) C_D = k^2 C_D\ .
\label{eq5.10}
\end{equation}
This equation can be solved numerically for a given tube, with the
usual phenomenology of bound and scattering states.

\section{Discussion}

It is helpful to interpret our results first in the nonrelativistic
limit.  We expect to find no coupling to the spin because the
nonrelativistic boundary conditions do not involve spin at all and the
exact Hamiltonian commutes with $\vec S$.  It is reassuring to see
that our formalism recovers this result: in the nonrelativistic limit,
$D\to 1$, so that in the parallel transport frame, the spin appears to
rotate with angular frequency $-\vec{\omega}$, where the minus sign
indicates that the sense of the rotation is opposite that of the
coordinate frame.  In other words, the spin vector maintains
its orientation in the three-dimensional Euclidean space in which the
tube is embedded (which corresponds to the coordinate frame of some
three-dimensional Euclidean experimenter).  Away from the
nonrelativistic limit, $D \ne 1$, so the variation of spin is no
longer exactly opposite to the rotation of the parallel transport
frame.

\begin{figure}[ht]
        $$\BoxedEPSF{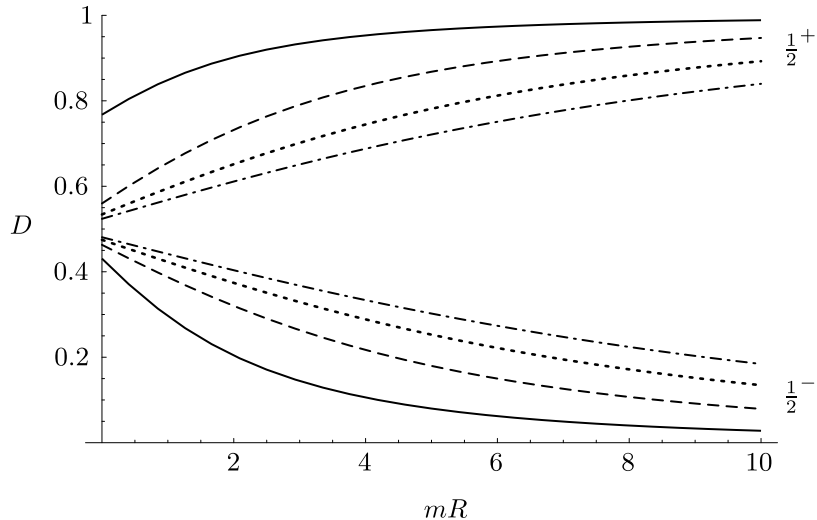}$$ \caption{Depolarizing factor $D$
        as a function of $m R$ for the $\frac{1}{2}^+$ and $\frac{1}{2}^-$
	towers of states in Fig.~\ref{fig1}. }  \label{fig3}
\end{figure}

Now if $D$ were to go to zero, the vector $\vec{J}$ would not rotate
relative to the axis of the tube.  This would parallel the phenomenon
described by Chiao and Wu.  However, $D$ is never zero in the lowest
energy subspace.  Even for $m=0$, $D \approx 0.767$.  We obtain a
complete description of $D$ as a function of $mR$ by solving for the
lowest root of the eigenvalue condition eq.~(\ref{eq4.5}) as a
function of $m$ and substituting in the expression eq.~(\ref{eq5.5})
for $D$.  A plot of $D$ versus $mR$ appears in
Fig.~(\ref{fig3}). The states with $\rho = \{\frac{1}{2}^+ , \lambda
\}$ do have $D\to 1$ as $m \to \infty$, as we expect.  However, the
states for which $\rho = \{\frac{1}{2}^- , \lambda \}$ have $D \to 0$
in the nonrelativistic limit.  This ``unphysical'' result appears as
an artifact of the adiabatic approximation, which breaks down for
these states in the nonrelativistic limit, as explained in Section
VI.

The effect of the factor $D$ on the vector $\vec{J}$ has a simple
geometric interpretation.  As an example, consider a tube whose
defining curve lies in a plane and such that it has two straight
sections, $|s| \geq s_0$, joined by a section with constant curvature
$\kappa$ and center of curvature $P$.  \footnote{Actually, we require
the transition from zero to nonzero curvature to occur smoothly over a
long enough interval that the adiabatic approximation is applicable.}
In addition, construct a complementary tube parameterized by the same
arc-length parameter $s$ satisfying the following conditions: it is
parallel to the first for $s \leq -s_0$; it has a section of constant
curvature $D \kappa$ with center of curvature $P$ for $|s| < s_0$; and
it has another straight section for $s \geq s_0$. The smaller
curvature implies a larger radius of curvature.  The effect of the
depolarizing factor can be summarized by the following construction:
as the Dirac particle traverses the curved section of the tube, the
vector $\vec{J}$ remains fixed relative to the tangent of the
complementary curve at the corresponding value of $s$.  For $m=0$, if
the tube curves through $\Delta \Theta = 2 \pi$, then $\vec{J}$ falls
behind the tangent by approximately $1.534\, \pi$.

In our formalism, we have used the boundary condition eq.~(\ref{eq1})
to simulate a hard-wall potential barrier.  Many of the qualitative
features of our results follow from the fact that spin is not a good
quantum number for the Dirac equation subject to such a boundary
condition.  It is natural to ask how sensitive these results are to
the specific choice of confining boundary condition.  An obvious
generalization given in eq.~(\ref{eq2}) is characterized by a chiral
angle $\theta$.  We have chosen $\theta = 0$ throughout; let us now
take $\theta \ne 0$.  If we set $\psi' =e^{i\gamma_{5}\theta/2}\psi$,
we can map this version back to the original boundary condition at the
cost of introducing a ``chiral'' mass term into the Dirac equation:
\begin{eqnarray}
\left(\vec{\alpha} \cdot \vec{p} + \beta m
e^{-i\gamma_{5}\theta}\right)\psi' &=& E \psi'  \\[2ex] -i \vec{\gamma}
\cdot
\hat{r} \psi' &=& \psi'\ .
\end{eqnarray}
Repeating the calculations in the earlier sections shows that the sole
effect of the chiral angle is to replace $m$ by $m \cos{\theta}$ in
the expression for the parameter $\zeta$ and thus implicitly in the
boundary conditions and depolarizing factor.  Thus the analysis above,
with an effective mass $m\cos\theta$, applies to this family of
problems.  In fact, for $\cos{\theta} \geq 0$, the plot in
Fig.~(\ref{fig2}) describes the depolarizing factor as a function of
$mR\cos{\theta}$.  When $mR\cos{\theta}$ becomes negative new
phenomena arise.  In particular, the boundary condition
eq.~(\ref{eq2}) acts like an effective attractive potential at $r=R$.
As a result for sufficiently negative $mR\cos{\theta}$ a qualitatively
new solution appears in each partial wave, which is ``bound to'' $r=R$
and decays exponentially as $r$ decreases.  The detailed study of
these states is beyond the scope of this paper.

In addition to these boundary conditions, there are other families of
boundary conditions that confine the probability current but which we
have not yet considered in detail.  Specifically one can
generalize eq.~(\ref{eq2}) by the condition
\begin{equation}
-i \vec{\gamma} \cdot \hat{n} \psi = e^{\Gamma} \psi
\end{equation}
where $\Gamma$ is any antihermitian Dirac matrix that anticommutes
with $\vec{\gamma} \cdot \hat{n}$.  $\Gamma$ need not be constant, as
the demonstration of probability confinement involves only algebraic
manipulations.  Interesting alternatives include
$\Gamma_{1}=\gamma_{5}\theta(s)$, $\Gamma_{2}=\vec\gamma\cdot \hat t$
and $\Gamma_{3}=\vec\gamma\cdot\hat t\times\hat r$.  Each of these
choices will generate different, nontrivial spin effects.  Thus we
believe that the essential feature of nontrivial spin transport should
not be unique to our choice of boundary condition.  Another direction
of possible future work is the generalization of our approach to
higher-dimensional spaces, as has been done with the nonrelativistic
theory.

\section{Acknowledgments}

R.L.J. would like to thank Dr.~Paolo Maraner for conversations
related to this work.  V.M.~and P.O.~thank the MIT Undergraduate
Research Opportunity Program (UROP) for support during the summer of
1998.


\begin{references}
\bibitem{TT} S.~Takagi and T.~Tanzawa, Prog.~Theor.~Phys., {\bf 87}
(1992) 561, and references therein.

\bibitem{NDIM} K.~Fujii, N.~Ogawa, Prog.~Theor.~Phys., {\bf 89}
(1993) 575; P.~Maraner, J.~Phys., {\bf A28} (1995) 2939,
Ann.~Phys.~(N.~Y.), {\bf 246} (1996) 325;  K.~Fujii, N.~Ogawa,
S.~Uchiyama, N.~M.~Chepilko, Int. J. Mod. Phys., {\bf A12} (1997)5235.

\bibitem{PM} P.~Maraner, hep-th/9809189.

\bibitem{OGAWA} For example, N.~Ogawa, Mod.~Phys.~Lett., {\bf A12}
(1997) 1583;  N.~Ogawa, hep-th/9703181;  N.M.~Chepilko and K.~Fujii,
Phys. At.  Nucl., {\bf 58} (1995) 1063.

\bibitem{CHIAO} R.Y.~Chiao and Y-S.~Wu, Phys.~Rev.~Lett., {\bf 57},
(1986) 933.

\bibitem{WignerEckart} E.P.~Wigner, {\it Gruppentheorie}, Friedrich
Vieweg und Sohn, Braunschweig, 1931;  C.~Eckart, Rev.~Mod.~Phys.,
{\bf 2} (1930) 305.

\bibitem{BAG} A.~Chodos, R.L.~Jaffe, K.~Johnson, C.B.~Thorn,
V.F.~Weisskopf,  Phys.~Rev.~{\bf D9} (1974) 3471, contains a
``physical'' derivation of this boundary condition in which the
particle is given a large mass outside the confining region.

\bibitem{Hannay} J.~Hannay, J.~Phys. {\bf A18} (1985) 221.

\bibitem{LIN} K.~Lin and R.L.~Jaffe, Phys. Rev. {\bf B54} (1996)
5750.
\end{references}
\end{document}